\documentclass[a4paper, 12pt, openany, oneside]{article}
\usepackage[cp1251]{inputenc}
\usepackage[english]{babel}
\usepackage{ graphics,amsfonts, amsmath,bm,amssymb, array}
\usepackage[dvips]{graphicx}
\usepackage[flushleft,small]{caption2}

\renewcommand{\@biblabel}[1]{#1.\hfill}

\newcommand{\diag}{\rm \diag\, }
\newcommand{\intl}{\int\limits}

\makeatother {

\begin{document}
\thispagestyle{empty} \large

\renewcommand{\refname}{\normalsize\begin{center}\bf
REFERENCES\end{center}}

\begin{center}
{\bf Interaction of the Electromagnetic p-Waves with Thin Metal Films}
\end{center}

\begin{center}
  \bf  A. V. Latyshev\footnote{$avlatyshev@mail.ru$} and
  A. A. Yushkanov\footnote{$yushkanov@inbox.ru$}
\end{center}\medskip

\begin{center}
{\it Faculty of Physics and Mathematics,\\ Moscow State Regional
University,  105005,\\ Moscow, Radio st., 10--A}
\end{center}\medskip

\begin{abstract}
For the first time it is shown that for thin metallic films
thickness of which not exceed thickness of skin -- layer, the
problem allows analytical solution for arbitrary boundary value
problems. The analysis of dependence of coefficients of
transmission, reflection and absorbtion on angle incidence,
thickness of films and coefficient of specular reflection is carried
out.
\medskip

\medskip

structures
\end{abstract}

\begin{center}\bf
  1. Introduction
\end{center}

The problem of interaction of an electromagnetic wave with the metal
films draws attention to itself for the long time \cite{F69} --
\cite{Landau8}. It is connected as with theoretical interest to this
problem, and with numerous practical appendices as well.

Researches of interaction of an electromagnetic wave with conducting
medium (in particular, metal films) were carried out basically for a
case of specular dissipation of electrons on a film surface. It is
connected with the fact that for more general boundary conditions
the problem becomes essentially complicated and does not suppose the
analytical solution generally. At the same time for the real
materials coefficient of specular reflection of electrons reflection
from surfaces is far from unit as a rule. For example, in the work
\cite{Zav} on the basis of the analysis of longitudinal magnetic
resistance of the thin metal wire it is shown, that for sodium
coefficient of specular reflection is equal to 0.3.

In the present work it is shown that for thin films, a thickness of
which does not exceed a thickness of a skin -- layer, the problem
allows the analytical solution for any boundary conditions.

Let's notice, that the most part of reasonings carrying out below is
fair for more general case of conducting medium  (in particular,
semi-conductor) films.\medskip

\begin{center}
{\bf  PROBLEM STATEMENT}
\end{center}

We consider the thin slab of conducting medium on which the
elec\-tro\-mag\-netic wave falls. The angle of incidence we denote
by $\theta$. We will assume, that the vector of magnetic field of
electromagnetic wave is parallel to the surface of slab. Such wave
is called $p$--wave \cite{K} or \cite{F69}.

We take Cartesian coordinate system with origin of coordinates on
one of the surfaces of a slab, with axes $x$, directed deep into the
slab. We direct the axes $y$ parallel with the vector of magnetic
field of electromagnetic wave.
At such choice of system of coordinates the 
electric field vector and magnetic field vector have the
following structure

$\mathbf{E}=\{E_x(x,y,z),0,E_z(x,y,z)\}$ and
$\mathbf{H}=\{0,H_y(x,y,z),0\}$.

Further components of electric and magnetic field vectors
we search in the form
$$
E_x(x,z,t)=E_x(x)e^{-i\omega t+ik\sin \theta z},\quad
E_z(x,z,t)=E_z(x)e^{-i\omega t+ik\sin \theta z},
$$
and
$$
H_y(x,z,t)=H_y(x)e^{-i\omega t+ik\sin \theta z}.
$$

Now behaviour of electric and magnetic fields of a wave in the slab
is described by the following system of differential equations
\cite{K}
$$
\left\{\begin{array}{l}
\dfrac{dE_z}{dx}-ik\sin\theta E_x+ikH_y=0 \\ \\
ikE_x-ik\sin\theta H_y=\dfrac{4\pi}{c}j_x\\ \\
\dfrac{dH_y}{dx}+ikE_z=\dfrac{4\pi}{c}j_z.
\end{array}\right.
\eqno{(1)}
$$

Here $c$ is the velocity of light, $\,\bf j$ is the density
of current, $k$ is the wave number.

We denote the thickness of slab by $d$.

The coefficients of transmission $T$, reflection $R$ and
absorption $A$ of the electromagnetic wave by slab are described
by following expressions
\cite{F69}, \cite{F66}
$$
T=\dfrac{1}{4}\big|P^{(1)}-P^{(2)}\big|^2,
\eqno{(2a)}
$$

$$
R=\dfrac{1}{4}\big|P^{(1)}+P^{(2)}\big|^2,
\eqno{(2b)}
$$
and
$$
A=1-T-R.
\eqno{(2c)}
$$

The quantities $P^{(j)}\;(j=1,2)$ are defined by following expressions
$$
P^{(j)}=\dfrac{\cos \theta+Z^{(j)}}{\cos \theta-Z^{(j)}},\qquad j=1,2.
\eqno{(3)}
$$

The quantities $Z^{(1)}$ and $Z^{(2)}$ correspond to impedance on
the lower surface of slab by symmetrical configuration of the
external magnetic field \big(this case 1 when
$$
H_y(0)=H_y(d),\qquad  E_x(0)=E_x(d),\qquad E_z(0)=-E_z(d)\;\big)
$$
and antysymmetrical configuration of the external magnetic field
(this case 2 when
$$
H_y(0)=-H_y(d),\qquad  E_x(0)=-E_x(d),\qquad
E_z(0)=E_z(d)\;\big).
$$

The impedance is thus defined as follows

$$
Z^{(j)}=\dfrac{E_z(-0)}{H_y(-0)}, \qquad j=1,2.
\eqno{(4)}
$$

We will consider the case when the width of the slab $d$ is less
than the depth of the skin -- layer $\delta$. Let's note, that depth
of the skin -- layer depends essentially on frequency of radiation,
monotonously decreasing in process of growth of the last. The value
$\delta$ possesses the minimum value in so-called infrared case
\cite{Landau10}
$$
\delta_0=\dfrac{c}{\omega_p},
$$
where $\omega_p$ is the plasma frequency.

For typical metals  \cite{Landau10} $\delta_0\sim 10^{-5}$ cm.

Hence for the films thickness of which $d$ is less than $\delta_0$
our assumption is fair for any frequencies.

The quantities $H_y $ and $E_z $ change a little on distances
smaller than depth of a skin -- layer.

Therefore under fulfilment of the given assumption $d<\delta$ these
electrical and magnetical fields will change a little in the slab.

In case 1 when $H_y(0)=H_y(d)$ it is possible to accept that the
value $H_y$ is constant within the slab.

Variation of the quantity of $y$-projection of electric field by
thickness of slab can be defined from the first equation of the
system (1)
$$
E_y(d)-E_y(0)=-ikd H_y+ik\sin\theta \intl_0^d E_x dx.
\eqno{(5)}
$$

From  the second equation of the system (1) it follows that on the
boundary of film the following relationship is satisfied
$$
E_x(0)=E_x(d)=H_y\sin\theta.
\eqno{(6)}
$$

The integral from relation (5) is proportional to the value of the
quantity normal to the surface component of electrical field on the
surface and therefore in according to the relation (6) to quantity
$H_y$.

We define the coefficient of proportionality as
$$
G=\dfrac{1}{{E_x(0)d}}\int\limits_{0}^{d}E_x(x)\,dx=
\dfrac{1}{H_y d\sin \theta}\int\limits_{0}^{d}E_x(x)\,dx.
\eqno{(7)}
$$

With the help of (7) we rewrite the relation (5) in the following
form
$$
E_z(d)-E_z(0)=\big(-ikd +ikGd\sin^2\theta\big)H_y.
$$

Considering antisymmetric character of the projection
of electric field $E_y$ in this case we receive
$$
E_y(0)=ikd\big(1-G\sin^2\theta\big)\dfrac{H_y}{2}.
\eqno{(8)}
$$

Therefore for the impedance we have
$$
Z^{(1)}=\dfrac{ikd}{2}\big(1 -G\sin^2\theta\big).
\eqno{(9)}
$$

For the case 2 when $E_z(0)=E_z(d)$, it is possible to assume that
$z$ -- projection of electric field $E_z$ is constant in the slab.

Then the magnetic field change on the width of a slab can be defined
from the third equation of the system (1)
$$
H_y(d)-H_y(0)=-ikdE_z+\dfrac{4\pi}{c}\intl_0^d j_z(x)dx.
\eqno{(10)}
$$

Thus
$$
j_z(x)=\sigma(x)E_z,
$$
where $\sigma(x)$ is the conductance that in general case depends
on coordinate $x$.

Let's introduce the longitudinal conductivity averaged by thickness
of the slab,

$$
\sigma_d=\dfrac{1}{E_zd}\intl_0^d j_z(x) dx=\dfrac{1}{d}
\intl_0^d \sigma(x) dx.
\eqno{(11)}
$$

Then the relation (10) can be rewritten in the following form
$$
H_y(d)-H_y(0)=-ikd E_z+\dfrac{4\pi d\sigma_d}{c}E_z.
$$

Considering sym\-met\-ry of the magnetic field, from here we have
$$
H_y(0)=\dfrac{1}{2}ikdE_z-\dfrac{2\pi d \sigma_d}{c}E_z.
$$

For the impedance (4) we have

$$
Z^{(2)}=\dfrac{2c}{ickd-4\pi \sigma_d d }.
\eqno{(12)}
$$

From here we receive expressions for the quantities
$P^{(j)}\;(j=1,2)$

$$
P^{(1)}=\dfrac{2\cos \theta+ikd(1-G\sin^2\theta)}
{2\cos \theta-ikd(1-G\sin^2\theta)},
\eqno{(13a)}
$$
$$
\quad P^{(2)}=\dfrac{(4\pi\sigma_d -ikc) d\cos \theta-2c}
{(4\pi\sigma_d -ikc) d\cos \theta+2c}.
\eqno{(13b)}
$$

We will assume that length of a wave of incident radiation surpasses
essentially the thickness of the slab, i.e. $kd\ll 1$. Then
expressions (9) and (12) for impedances and expression (13) for
quantity $P^{(j)}\;(j=1,2)$ become a little simpler
$$
Z^{(1)}_0=-\dfrac{1}{2}ikGd\sin^2\theta,
 \qquad Z^{(2)}_0=-\dfrac{c}{2\pi \sigma_d d }.
\eqno{(14)}
$$
Substituting (14) into (3), we have:
$$
P^{(1)}_0=
\dfrac{2\cos \theta-ikGd\sin^2\theta}{2\cos \theta+ikGd\sin^2\theta},
\qquad
P^{(2)}_0=
\dfrac{2\pi \sigma_d d\cos \theta-c}{2\pi \sigma_d d\cos \theta+c}.
\eqno{(15)}
$$

Thus quantities $R, T, A $ can be found according to the formulas
(2).

In a limiting case of non-conducting slab, when $\sigma_d\to 0$,
$G\to 1$ from these expressions we have
$$
P^{(1)}=-P^{(2)}=\dfrac{2+ikd\cos \theta}{2-ikd\cos \theta},
$$
from which
$$
T= 1,\qquad \, R= 0, \,\qquad A=0.
$$

Under almost tangent incidence when $\theta\to \pi/2$ we receive
$P^{(1)}\to -1,P^{(2)}\to -1$. Thus, we obtain that $T\to 0,\, R\to
1,\,A\to 0$.

Let the relation $kl\ll 1$ be true. Then in a low-frequency case,
when $\omega\to 0$, the quantity $\sigma_d$ for a metal film can be
presented in the following form \cite {S}

$$
\sigma_d=\dfrac{w}{\Phi(w)}\,\sigma_0,\quad\quad
w=\dfrac{d}{l},
\eqno{(16)}
$$
where
$$
\dfrac{1}{\Phi(w)}=\dfrac{1}{w}-\dfrac{3}{2w^2}(1-p)\intl_1^\infty\Big(
\dfrac{1}{t^3}-\dfrac{1}{t^5}\Big)\dfrac{1-e^{-wt}}{1-pe^{-wt}}dt.
$$

Here $l$ is the mean free path of electrons,
$p$ is the coefficient of specular reflection,
$\sigma_0=\omega_p^2\tau/(4\pi)$ is the static conductivity of
volume pattern, $\tau=l/v_F$ is the time of mean free path of
electrons, $v_F$ is the Fermi's velocity.

In a low-frequency case when the formula (16) is applicable, the
coefficients $T, R, A $ according to formulas (2) do not depend on
frequency of the incoming radiation.

For any frequencies these expressions are true under condition, that
it is necessary to use the following expression $ l\to \dfrac{v_F
\tau}{1-i\omega\tau}, $ as a quantity $l$ and the expression $
\sigma_0\to \dfrac{\sigma_0}{1-i\omega\tau}$ instead of $\sigma_0$.

For a case $kl\ll 1$ the quantity $G$ can be calculated from the
problem of behaviour of a plasma slab in variable electric field,
which is perpendicular to surface of slab \cite {LY2008}.

Let's calculate the coefficients of transmission and reflection in
the case when $kd\ll 1$. We will substitute expressions $P_0^{(1)}$
and $P_0^{(2)}$, defined according to (15), into the formulas (2).
We receive
$$
T=\cos^2\theta\Bigg|\dfrac{1-ik\dfrac{d}{2}G\sin^2\theta
\dfrac{2\pi d\sigma_d}{c}}{(\cos\theta+ik\dfrac{d}{2}G
\sin^2\theta)(1+\dfrac{2\pi d\sigma_d}{c}\cos\theta)}\Bigg|^2,
\eqno{(17)}
$$
$$
R=\Bigg|\dfrac{ik\dfrac{d}{2}G\sin^2\theta-
\dfrac{2\pi d\sigma_d}{c}\cos^2\theta}{(\cos\theta+ik\dfrac{d}{2}G
\sin^2\theta)(1+\dfrac{2\pi d\sigma_d}{c}\cos\theta)}\Bigg|^2,
\eqno{(18)}
$$

Let's consider a case of a thin slab of sodium. Thus \cite{F69}
$\omega_p=6.5\times 10^{15} sec^{-1}$, $v_F=8.52\times 10^7$ cm/sec. We
consider a case of low frequencies $0<\omega \leqslant 0.2\omega_p$.
In this case the quantity  $G<1$, therefore $|kdG|\ll 1$. Formulas
(17) and (18) become simpler essentially and have the following form
$$
T=\dfrac{1}{\Big|1+\dfrac{2\pi d\sigma_d}{c}\cos \theta\Big|^2}
\eqno{(19)}
$$
and
$$
R=\Bigg|\dfrac{\dfrac{2\pi d\sigma_d}{c}\cos \theta}
{1+\dfrac{2\pi d\sigma_d}{c}\cos \theta}\Bigg|^2.
\eqno{(20)}
$$

Using formulas (19) and (20), we will carry out graphic research
of coefficients of transmission, reflection and absorption.

\begin{center}
 \bf DISCUSSION OF RESULTS
\end{center}

Let's put, that $ \tau\omega_p=0.001$ \cite {F69}.
Then $\tau=1.5\cdot 10^{-13} $ sec,
i.e. $\nu =\dfrac{1}{\tau}=0.667\cdot 10^{13}$  sec$^{-1}$.

The coefficient of specular reflectivity we will take equal to $0.3$
(Figs. 1--6 ).

We investigate dependence of coefficients of transmission,
reflection and absorption on quantity of oscillations frequency of
electromagnetic field ($0 <\omega \leqslant 0.2\omega_p $).

On figs 1 - 3 we will represent dependence of coefficients of
transmission (fig.1 ), reflection (fig. 1) and absorption (fig. 3)
on varying values of quantities of thickness of the slab in case of
normal incidence of the electromagnetic wave ($ \theta=0$). The
curves $1,2,3$ correspond to the following values of thickness of
slab $d=10^{-6}$ cm, $0.9\cdot 10^{-6}$ cm, $0.8\cdot 10^{-6}$ cm.

Plots in these figures show, that in the whole range of change of
frequency of electromagnetic wave oscillations $0 \leqslant \omega
\leqslant 0.2\omega_p $ values of coefficients of transmission and
absorption decrease with growth of a thickness of the slab, and
values of coefficients of reflection increase.

Thus the absorption coefficients has
maximum at $\omega\approx 10^{14}$ sec$^{-1}$.

On figs 4--6 we represent dependence of coefficients of
transmission (fig. 4), reflection (fig. 5) and absorption (fig. 6)
at various values quantities of  angle of falling
of an electromagnetic wave.

Graphics in these figures show that in all range of change of
frequency of electromagnetic wave oscillations with growth of
quantity an angle of incidence of the electromagnetic wave values of
transmission and absorption increase, and values of reflection
coefficient decreases. At small values of an angle of incidence
change of the absorption coefficient have non-monotonic character.

On figs 7 -- 9 dependence of coefficients of transmission (fig. 7),
reflection (fig. 8) and absorption (fig. 9) on various values of
coefficients of specular reflection is represented. Curves $1,2,3$
correspond to values of coefficients of specular reflection $p=0,
0.5, 0.8$. The thickness of the slab is equal to $d=10^{-6}$ cm.

The analysis of these figures shows, that in whole range of change
frequencies of oscillations of an electromagnetic wave values of
transmission and absorption coefficients decrease, and values of
coefficients of specular reflection increase with growth of
coefficient of specular reflection. Under small values of
coefficients of specular reflection change of absorption
coefficients has non-monotonic character.

\begin{figure}[ht]\center
\includegraphics[width=14.0cm, height=14cm]{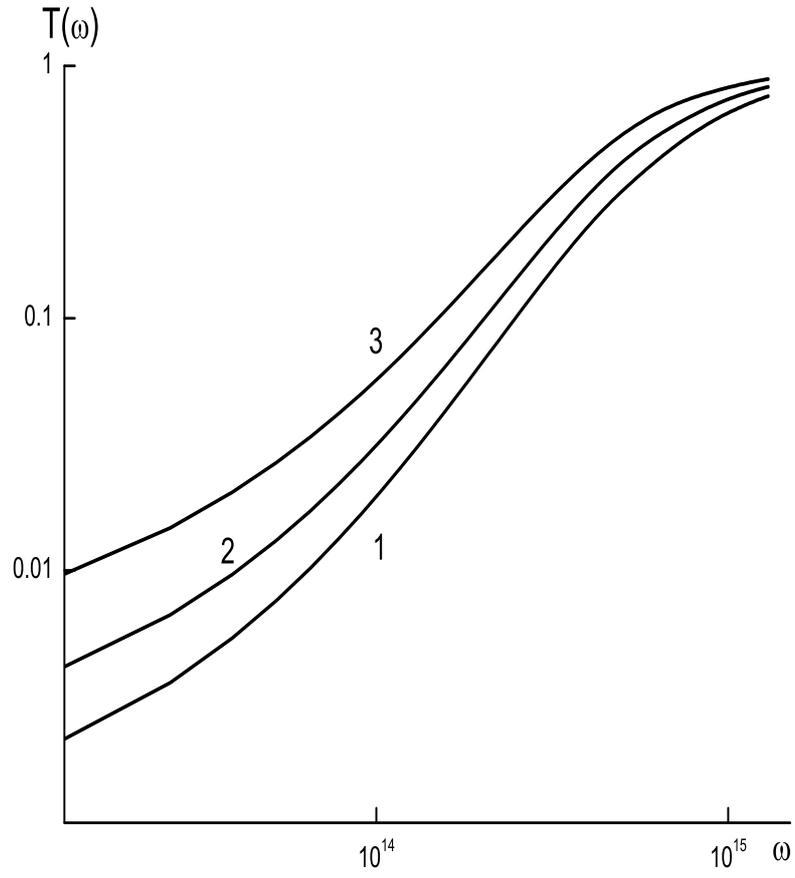}
\noindent\caption{Transmission coefficient. Curves $1,2,3$
correspond to quantities the thickness of slab $d=10^{-6}$ cm,
$0.9\cdot 10^{-6}$ cm, $0.8\cdot 10^{-6}$ cm.
}\label{rateIII}
\end{figure}

\begin{figure}[hb]\center
\includegraphics[width=14.0cm, height=9cm]{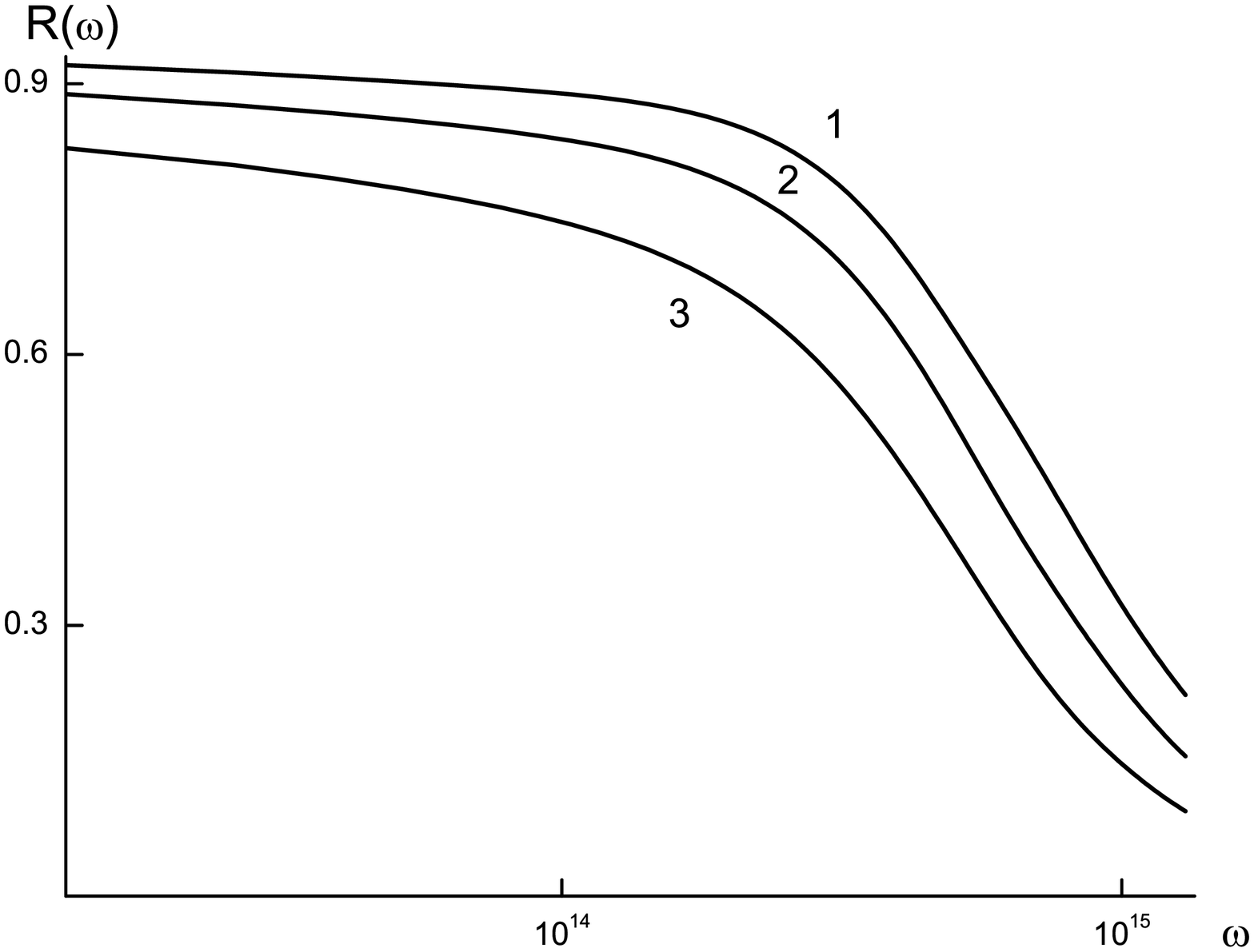}
\noindent\caption{Reflection coefficient.
Curves $1,2,3$
correspond to quantities the thickness of slab $d=10^{-6}$ cm,
$0.9\cdot 10^{-6}$ cm, $0.8\cdot 10^{-6}$ cm.
}\label{rateIII}
\end{figure}

\begin{figure}[hb]\center
\includegraphics[width=14.0cm, height=9cm]{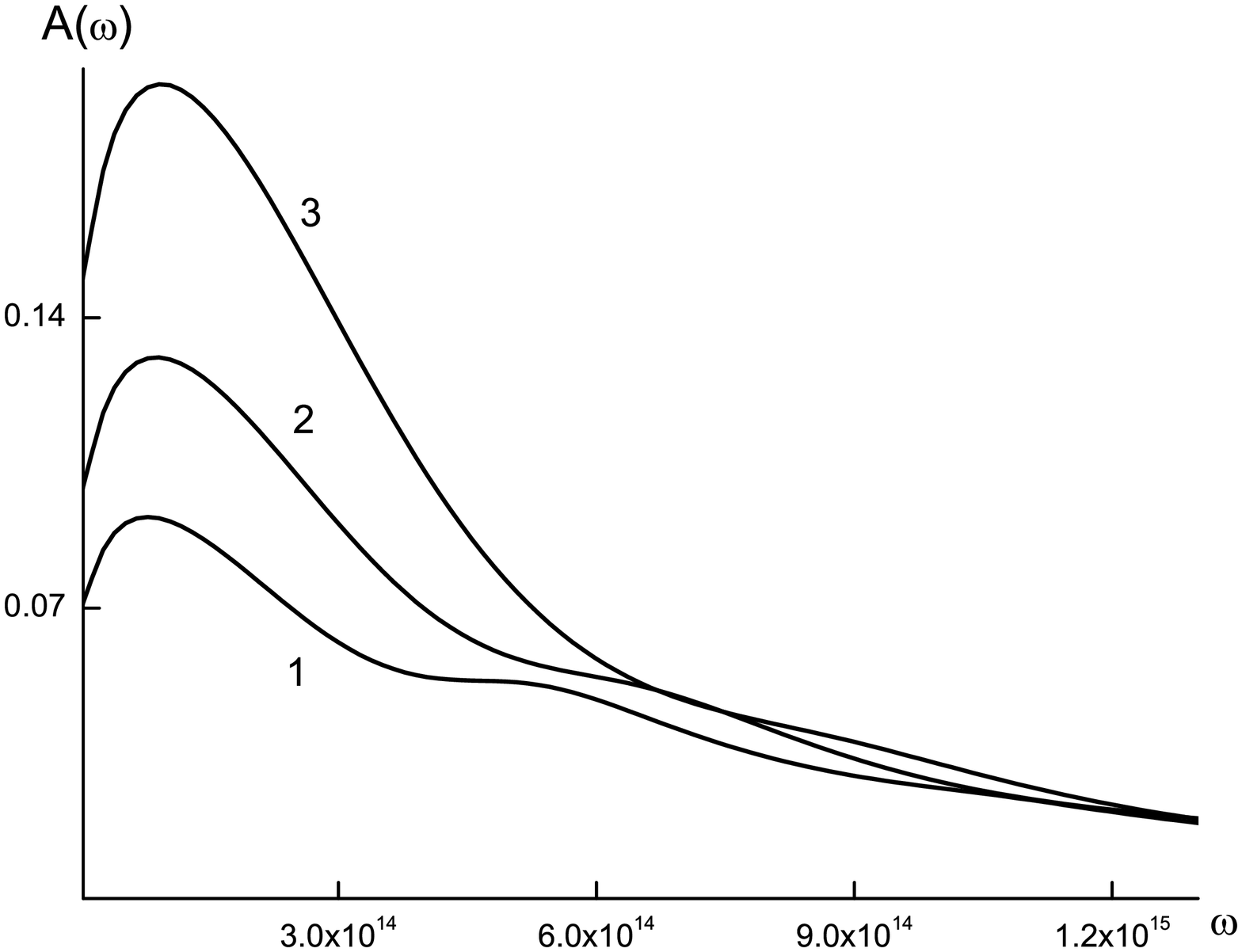}
\noindent\caption{Absorption coefficient. Curves $1,2,3$
correspond to quantities the thickness of slab $d=10^{-6}$ cm,
$0.9\cdot 10^{-6}$ cm, $0.8\cdot 10^{-6}$ cm.
}\label{rateIII}
\end{figure}

\begin{figure}[ht]\center
\includegraphics[width=14.0cm, height=9cm]{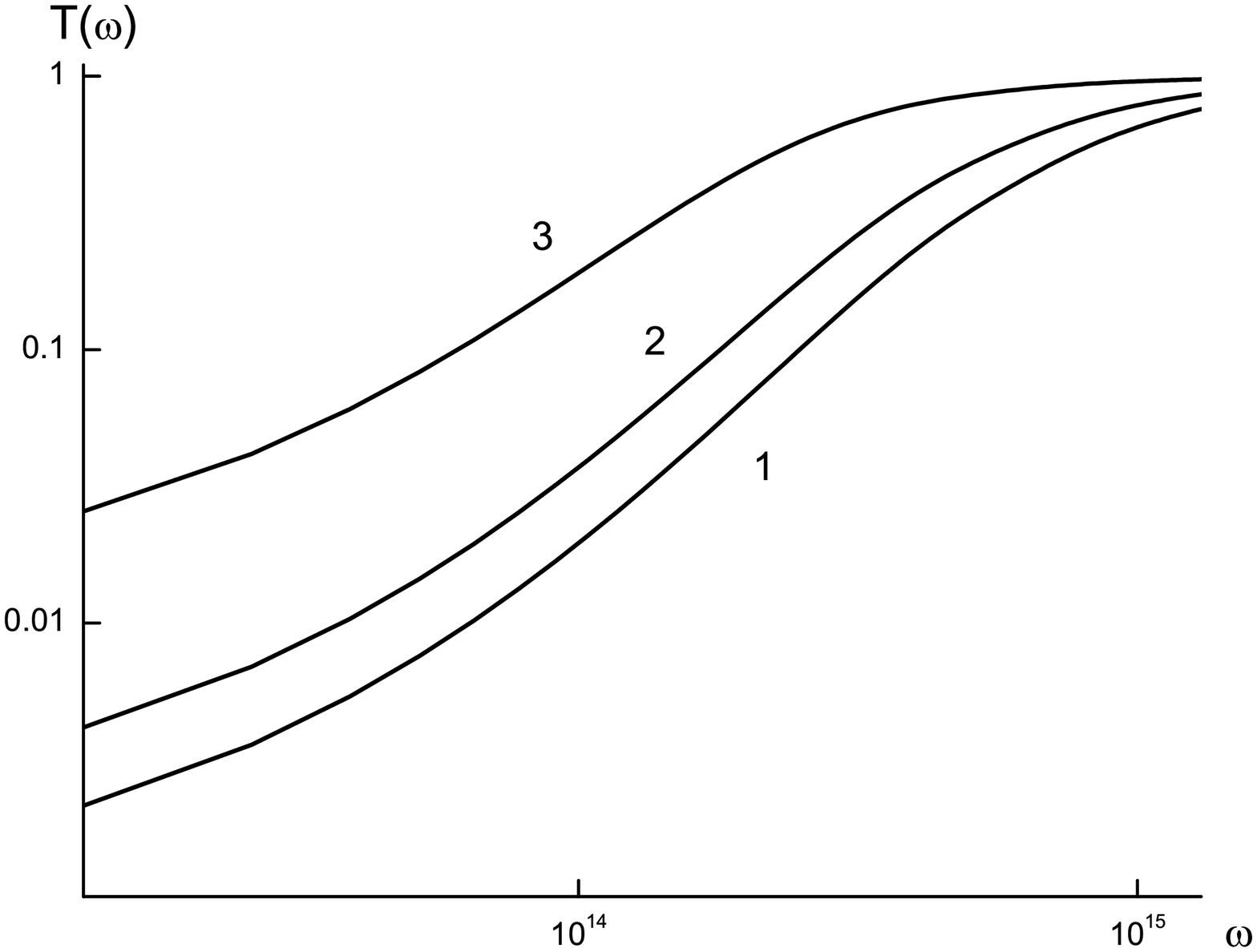}
\noindent\caption{Transmission coefficient.
Curves $1,2,3$ correspond to quantities the angle of incidence
$\theta=0, \dfrac{\pi}{4}, \dfrac{5\pi}{12}$.
}\label{rateIII}
\end{figure}

\begin{figure}[hb]\center
\includegraphics[width=14.0cm, height=9cm]{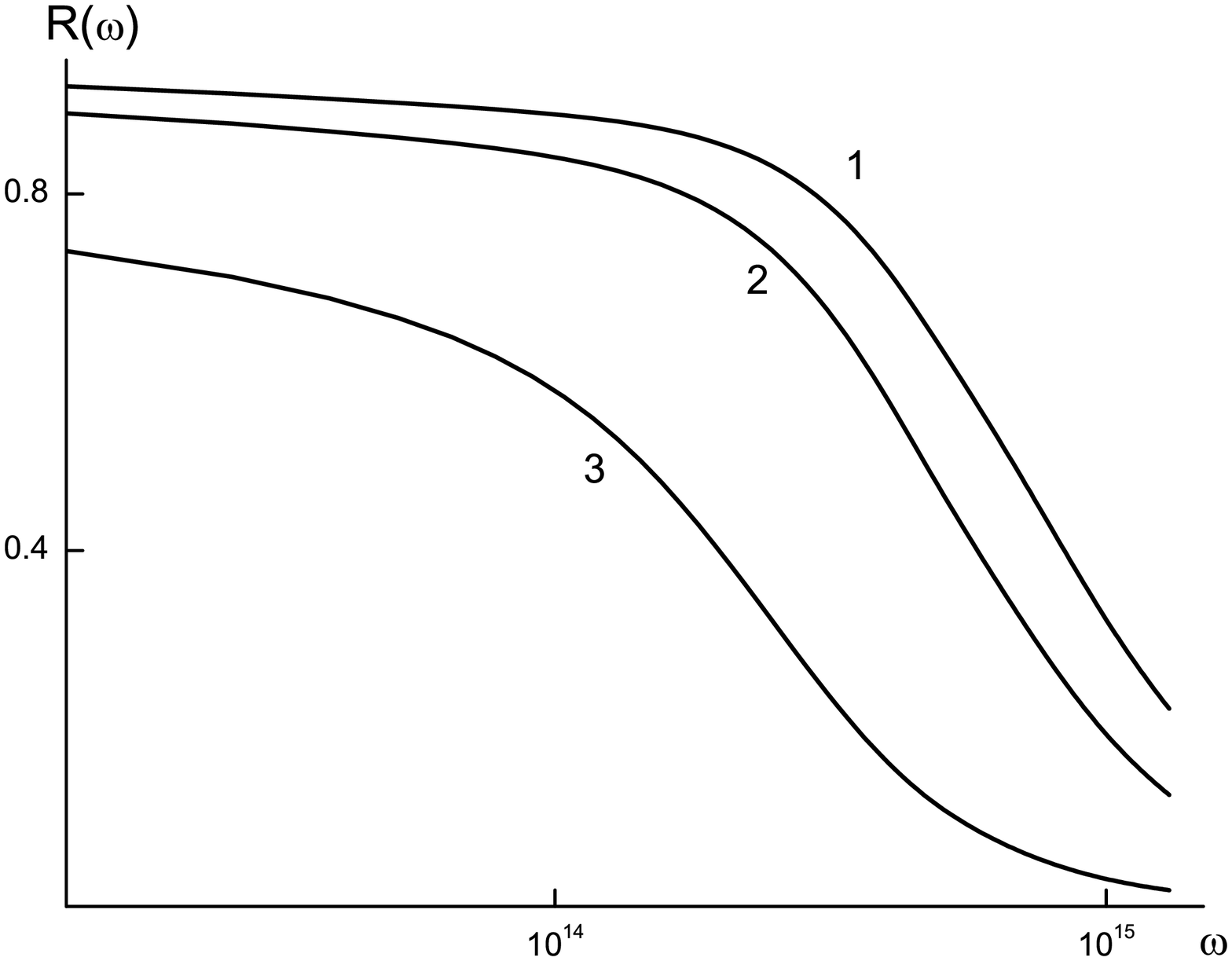}
\noindent\caption{Reflection coefficient.
Curves $1,2,3$ correspond to quantities the angle of incidence
$\theta=0, \dfrac{\pi}{4}, \dfrac{5\pi}{12}$.
}\label{rateIII}
\end{figure}

\begin{figure}[hb]\center
\includegraphics[width=14.0cm, height=9cm]{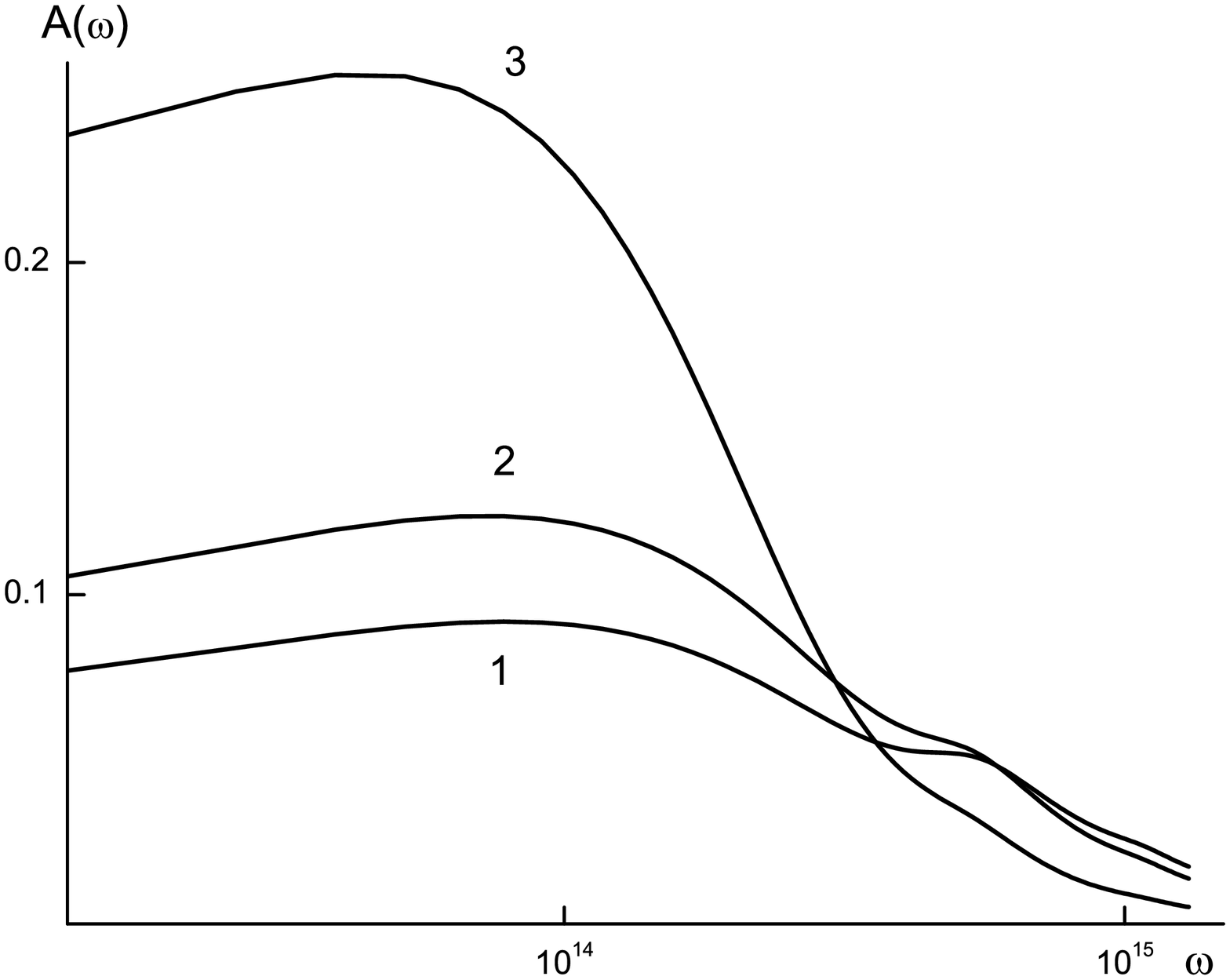}
\noindent\caption{Absorption coefficient.
Curves $1,2,3$ correspond to quantities the angle of incidence
$\theta=0, \dfrac{\pi}{4}, \dfrac{5\pi}{12}$.
}\label{rateIII}
\end{figure}

\begin{figure}[ht]\center
\includegraphics[width=14.0cm, height=9cm]{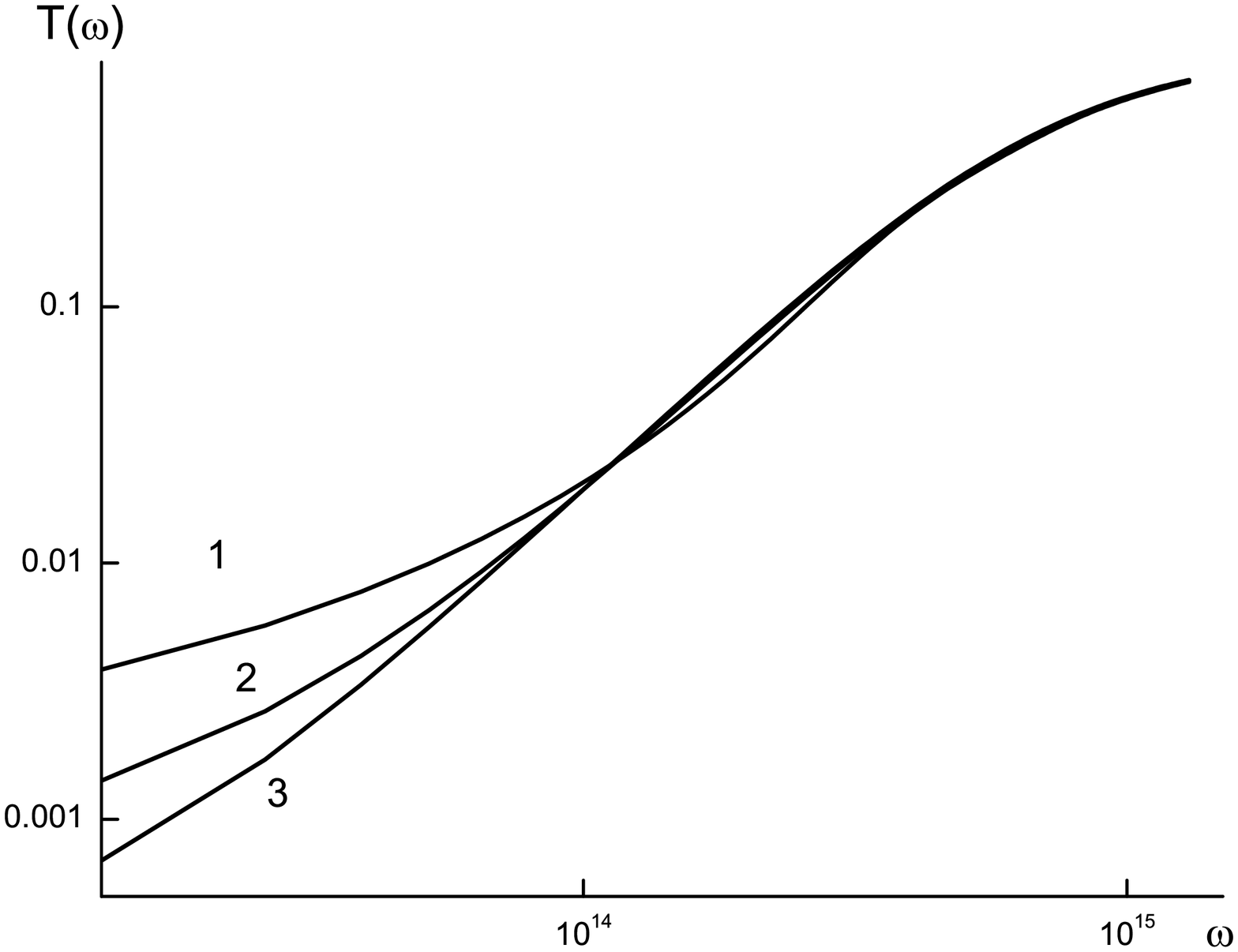}
\noindent\caption{Transmission coefficient.
Curves $1,2,3$ correspond to quantities the coefficient of
specular reflection $p=0, 0.5, 0.8$.
}\label{rateIII}
\end{figure}

\begin{figure}[hb]\center
\includegraphics[width=14.0cm, height=9cm]{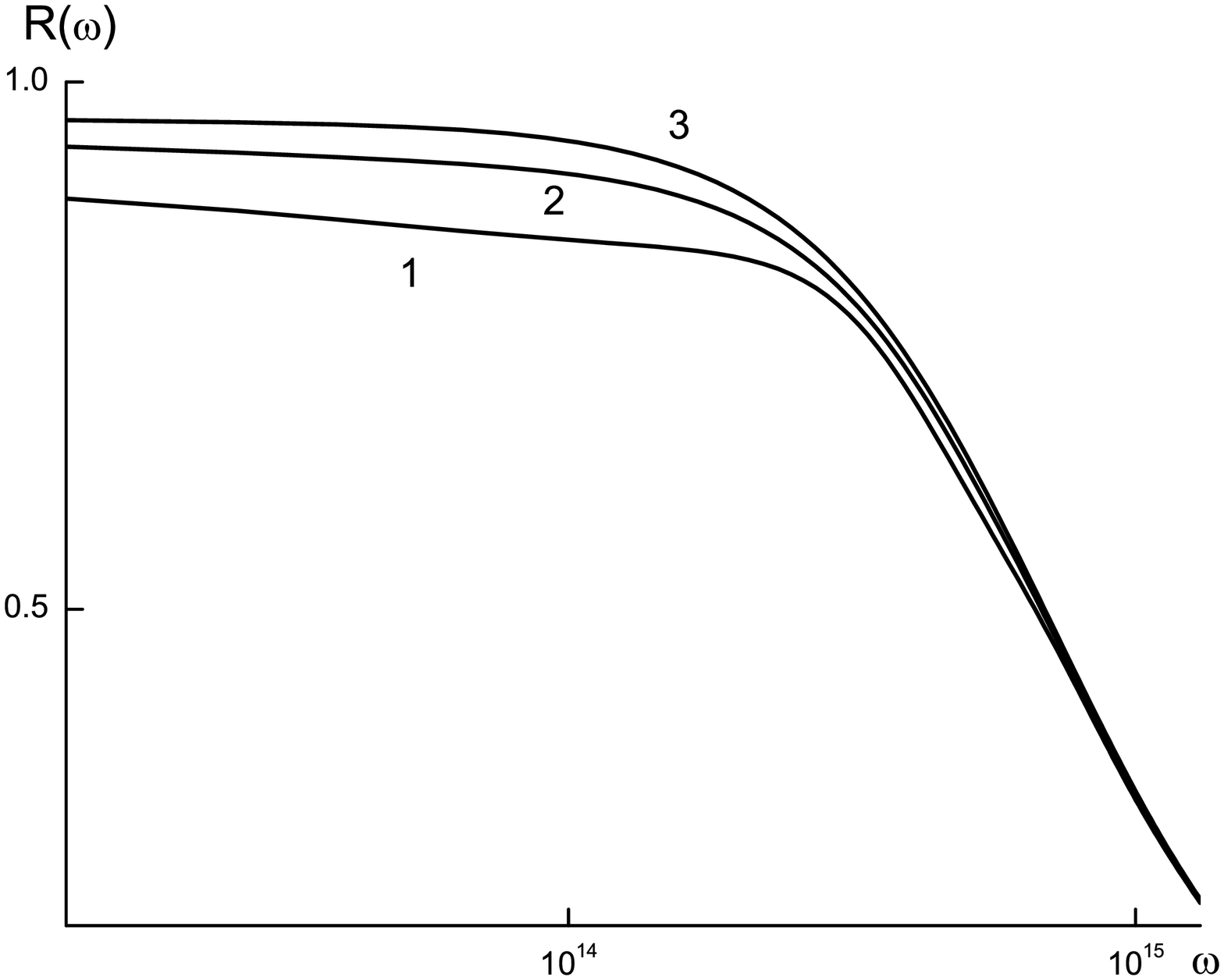}
\noindent\caption{Reflection coefficient.
Curves $1,2,3$ correspond to quantities the coefficient of
specular reflection $p=0, 0.5, 0.8$.
}\label{rateIII}
\end{figure}

\begin{figure}[hb]\center
\includegraphics[width=14.0cm, height=9cm]{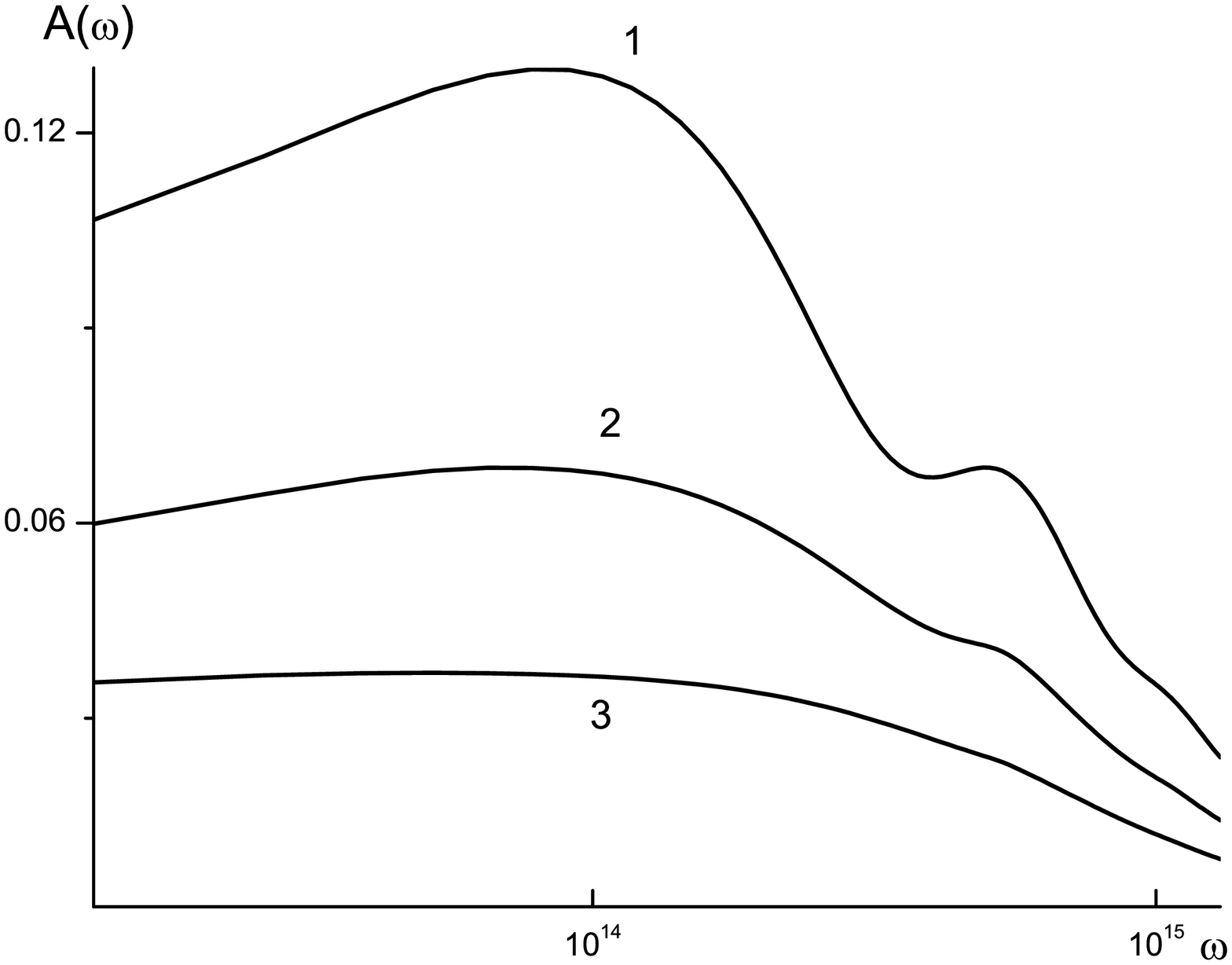}
\noindent\caption{Absorption coefficient.
Curves $1,2,3$ correspond to quantities the coefficient of
specular reflection $p=0, 0.5, 0.8$.
}\label{rateIII}
\end{figure}

\end{document}